\documentclass[prb,twocolumn,showpacs]{revtex4}
\input epsf

\usepackage{graphics}
\usepackage{amsmath}
\usepackage{bm}

\newcommand{\Journal}[4]{#1 \textbf{#2}, #3 (#4)}
\newcommand{\PRev}{ Phys. Rev. }
\newcommand{\JPSJ}{ J. Phys. Soc. Jpn.}
\newcommand{\be}{\begin{eqnarray}}
\newcommand{\ee}{\end{eqnarray}}
\newcommand{\nn}{\nonumber}

\newcommand{\pr}{PrOs$_4$Sb$_{12}$}

\begin{document}
\title{Symmetry properties of  the nodal superconductor \pr}
\author{T. R. Abu Alrub}
\author{S. H. Curnoe}
\affiliation{Department of Physics and Physical Oceanography,
Memorial University of Newfoundland, St. John's, NL, A1B 3X7, Canada}
\begin{abstract}
We present a theoretical study of the 
superconducting gap function
in \pr\ using a symmetry-based approach.
A three-component order parameter in the triplet channel best 
describes superconductivity.
The gap function is non-degenerate
and the lower branch has four cusp nodes at unusual points of the Fermi surface, which lead
to power law behaviours  in the density of states, specific heat
and nuclear spin relaxation rate.  

\end{abstract}


\maketitle
\section{Introduction}
By most accounts, \pr\ is an unconventional 
superconductor.\cite{Maple2001,Bauer2002,Izawa2003,Aoki2003,Chia2003,Huxley2004,Frederick2005,Nishiyama2005,
Higemoto2007} 
The superconducting phase breaks time-reversal
symmetry\cite{Aoki2003} and the paired electrons are in a spin triplet 
configuration.\cite{Higemoto2007}
The existence of point nodes in the superconducting gap function
is indicated by power law behaviour in the temperature
dependencies of specific heat,\cite{Bauer2002,Frederick2005}
penetration depth,\cite{Chia2003} 
thermal conductivity\cite{Izawa2003}, and Sb-NQR; 
\cite{Katayama2007} however other experiments find the gap function
to be nodeless.\cite{MacLaughlin2002,Kotegawa2003,Suderow2004}
Two distinct features in the specific heat\cite{Aoki2002,Aoki2003,Vollmer2003} and other measurements\cite{Chia2003,Izawa2003,Tayama2003,Ho2003,
Oeschler2004,Grube2006}
were initially interpreted as two phase transitions involving a change
in symmetry of the superconducting order parameter, but recently
these results have been ascribed to sample inhomogeneity or
two-band superconductivity.\cite{Measson2004,Seyfarth2005,Seyfarth2006,Yogi2006}
On the theoretical side, several phenomenological unconventional 
order parameters
have been proposed\cite{Goryo2003,Ichioka2003,Miyake2003,Maki2004} and unconventional pairing 
mechanisms have been studied.\cite{Matsumoto2005,Thalmeier2006}
In light of all these intriguing and somewhat contradictory findings,
it is not surprising that the only consensus on the symmetry of the
superconducting order parameter is that it is
probably unconventional.

In this paper, we will consider the 
results of a strict analysis of symmetry and symmetry-breaking
described by Landau theory.\cite{Volovik,Sigrist1991,Sergienko2004}
According to this approach, the order parameter which describes the
normal to superconducting phase transition must belong to one of the
irreducible representations of the crystallographic point group. 
Each irreducible representation yields a limited number of 
superconducting phases.  The most 
convenient and accurate way to label the various phases is by 
their symmetry groups.   
All of the superconducting 
symmetry groups are subgroups of the normal phase
symmetry $G\times U \times 
{\cal K}$, where $G$
is the point group of the crystal, $U$ is 
$U(1)$ gauge (phase) symmetry and ${\cal K}$ is time-reversal. 
Some of the subgroups include elements which are 
non-trivial combinations of phases, time reversal and point group elements.
As described by Sigrist and Ueda\cite{Sigrist1991} and Volovik and
Gor'kov,\cite{Volovik} strong spin-orbit coupling is assumed in this
classification scheme.

The point group symmetry of \pr\ is $T_h$ (tetrahedral), 
which has a one-dimensional
representation $A_{g,u}$, a two-dimensional representation $E_{g,u}$ and a three-dimensional
representation $T_{g,u}$, in each of the singlet (subscripted by $g$) and triplet
(subscripted by $u$) channels.
The $A_g$ order parameter describes a ``conventional" or ``$s$-wave"
superconductor.  It is associated with a single, fully gapped
superconducting phase.  The $A_u$ order parameter describes
triplet superconductivity, also with a single, fully gapped 
superconducting phase.  The 
$A_{g,u}$ phases have symmetry $T\times {\cal K}$, where $T$ is the tetrahedral point group.
The  $E_{g,u}$ and $T_{g,u}$ order parameters are each associated
with more than one superconducting phases, corresponding to different
symmetries.  
The 
$E_{g,u}$ order parameters describe three different superconducting phases, of which
two are accessible from the normal state via a second order phase transition,
while the $T_{g,u}$ order parameters describe nine different superconducting phases,
of which four are accessible from the normal state.
The symmetry properties of all of these states and their corresponding gap
nodes are given in Table I of Ref.~\onlinecite{Sergienko2004}.\cite{error}

The order parameter which best describes experiments is $T_u$, the three
component order parameter in the triplet channel.
Broken time reversal symmetry rules out the $A_{g,u}$ order parameters.
The $E_{g,u}$ phase that is accessible from the normal phase and 
that breaks time reversal symmetry is $T(D_2)$,
which has point nodes in the
$\langle111\rangle$ directions which are not indicated in any experiment.  
The $T_g$ phases which are accessible from the normal state
have either time reversal symmetry, line nodes, or nodes in the 
$\langle111\rangle$ directions, leaving $T_u$ as the only possibility.
There are two $T_u$ phases accessible from the normal phase
that break time reversal symmetry: $C_3(E)$ and
$D_2(E)$; the former
has nodes in the $\langle 111\rangle$ directions, 
leaving the phase $D_2(E)$ 
as the most likely candidate.
The elements of the symmetry group $D_2(E)$ are
$\{E,C_2^x{\cal K},
U_1(\pi)C_2^y{\cal K},U_1(\pi)C_2^z\}$, where $E$ is the identity,
$U_1(\pi)$ are phases, $C_2^i$ are rotations of $\pi$ about the 
$i$-axis, and  ${\cal K}$ is time
reversal. The triplet $D_2(E)$ phase has four point nodes in the
$[\pm \alpha, \pm \beta,0]$ directions.
The proof that $D_2(E)$ has nodes in the triplet channel
is given in the Appendix.

The issue of whether there are two different superconducting phases
(as suggested by specific heat and 
thermal conductivity experiments\cite{Aoki2003,Aoki2002,Izawa2003,Ho2003,Vollmer2003,Tayama2003}) or only one 
(according to the two-band superconductivity scenario\cite{Measson2004,Seyfarth2005,Seyfarth2006,Yogi2006}) is to some
extent by-passed by a fluke of Landau theory:  the $D_2(E)$ phase is accessible 
via second order phase transitions both
directly from the normal phase and via an intermediate phase 
$D_2(C_2)\times {\cal K}$.
Thus it is a viable candidate for either situation.
Therefore, we identify $D_2(C_2)\times {\cal K}$ as the `A-phase' and
$D_2(E)$ as the `B-phase', and we will consider 
both the case when the A-phase is present and the case when the A-phase is
absent on the
phase diagram.
Note that the elements of the group  $D_2(C_2)\times{\cal K}$ are $\{E,C_2^x,
U(\pi)C_2^y,U(\pi)C_{2}^z\}\times{\cal K}$ and that $D_2(E)$ is a subgroup of
$D_2(C_2)\times{\cal K}$.

Recently, microscopic weak coupling theory has been applied to
tetrahedral superconductors\cite{Kuz2005,Mukherjee2006}, and it was
shown that the phase $D_2(C_2)\times{\cal K}$ is stable, while
$D_2(E)$ is not.\cite{Mukherjee2006}
This is apparently in disagreement with the observation of
broken time reversal symmetry, which means either that
\pr\ is a strong coupling superconductor,
as claimed in Refs.\ \onlinecite{Vollmer2003,Kotegawa2003,Grube2006,Seyfarth2006} or that the B-phase is better
described as a $D_2(C_2)\times{\cal K}$ phase.  We shall not pursue this possibility here,
apart from noting that there are still issues whose resolution may change
the conclusions of this work.

\section{The Superconducting Gap Function}

The superconducting gap function is a $2\times 2$ matrix in pseudospin space,
\begin{equation}
{\widetilde \Delta}({\bm k}) = i\tilde\sigma_y \psi({\bm k})  = \left( \begin{array}{cc}
  0 & \psi({\bm k}) \\
  -\psi({\bm k}) & 0 
\end{array} \right)
\end{equation}
in the singlet channel, and
\begin{equation}
{\widetilde \Delta}({\bm k}) = i[{\tilde{\bm\sigma}}\cdot{\bm d}({\bm k})]
\tilde\sigma_y = 
 \left( \begin{array}{cc}
-d_x({\bm k}) + id_y({\bm k}) & d_z({\bm k})\\
d_z({\bm k}) & d_x({\bm k}) + id_y({\bm k}) 
\end{array} \right)
\end{equation}
in the triplet channel, where $\psi({\bm k})$ and ${\bm d}({\bm k})$
are even and odd functions of ${\bm k}$, respectively.
For singlet pairing, the gap function is given by
\begin{equation}
\Delta({\bm k}) = |\psi({\bm k})|,
\end{equation}
while for triplet pairing the gap function may be non-degenerate,
\begin{equation}
\Delta_{\pm}({\bm k}) = 
\left[|{\bm d}({\bm k})|^2 \pm |{\bm q}({\bm k})|\right]^{1/2},
\label{gap}
\end{equation}
where ${\bm q}({\bm k}) = i{\bm d}({\bm k}) \times {\bm d}^{*}({\bm k})$.
When ${\bm d}({\bm k})$ is real ${\bm q}({\bm k})$ vanishes and the gaps are degenerate and unitary.
Otherwise, the gap is non-degenerate and the lowest energy branch has a cusp where the two branches meet.

The gap function may be expanded in terms of the basis functions for a 
single representation of the point group,
\begin{eqnarray}
\psi({\bm k}) &=& \sum_i\eta_i\psi_i({\bm k}) \\
{\bm d}({\bm k})& =& \sum_i\eta_i{\bm d}_i({\bm k})
\label{d(k)}
\end{eqnarray}
where $\psi_i({\bm k})$ and ${\bm d}_i({\bm k})$ are basis functions for
even (spin-singlet) and odd (spin-triplet) 
representations of the point group, respectively, and
$\eta_i$ are components of the order parameter.  For the remainder of
this article we will limit our discussion to the three component order
parameter in the triplet channel $T_u$.  An appropriate set of
basis functions for this representation are\cite{Sergienko2004}
\begin{eqnarray}
\bm{d_1} &\sim& a k_y \hat{\bm z} + b k_z \hat{\bm y},\nn\\
\bm{d_2} &\sim& a k_z \hat{\bm x} + b k_x \hat{\bm z},\nn\\
\bm{d_3} &\sim& a k_x \hat{\bm y} + b k_y \hat{\bm x}.
\label{basis}
\end{eqnarray}
where $a$ and $b$ are arbitrary real numbers.  More general forms,
which include higher orders in ${\bm k}$, are considered in the Appendix.

The phases associated with each representation are minima of the Landau
potential, which is expanded in terms of the order parameter.
The transformation properties of the basis functions
(\ref{basis}) get transferred to the order parameter, and the Landau
potential is constructed to be invariant under all operations
of the space group, gauge transformations and time reversal. 
The Landau potential also determines which phases are connected by 
second order phase transitions.  A complete analysis of the Landau potentials
for the tetrahedral point group $T$ is given in Ref.~\onlinecite{Sergienko2004}.
The three component order parameter $(\eta_1,\eta_2,\eta_3)$, 
defined by (\ref{d(k)}) and (\ref{basis}),
has four phases which are accessible from the normal state by a second
order phase transition, $(0,0,1)$, $(1,1,1)$, $(1,e^{2\pi i/3},e^{-2\pi i/3})$ and
$(0,i|\eta_2|,|\eta_1|)$, with symmetries $D_2(C_2)\times {\cal K}$,
$C_3\times {\cal K}$, $C_3(E)$ and $D_2(E)$ respectively.
Thus the components of the order parameter in the A-phase are
 $(0,0,1)$ (or, more precisely, $(0,0,|\eta_1|)$)
and in the B-phase are $(0,i|\eta_2|,|\eta_1|)$.  These
statements are summarised in Table I.  Different
domains of each phase are obtained by permuting the components;
the analysis below uses this particular choice of domain.  A discussion 
of domains appears in Section \ref{sec-domains}.

\begin{table}[ht]
\begin{tabular}{l|rlcll}
\hline
phase & normal & $\rightarrow$ & A & $\rightarrow$ & B\\
OP components &$(0,0,0)$ & $\rightarrow$ & $(0,0,|\eta_1|)$ &
$\rightarrow$& $(0,i|\eta_2|,|\eta_1|)$ \\
symmetry group & $T_h\times U \times {\cal K}$ & $\rightarrow$ & 
$D_2(C_2)\times{\cal K}$ & $\rightarrow$ & $D_2(E)$ 
\\
\hline
\end{tabular}
\caption{Order parameter (OP) components and symmetry group elements
for the proposed normal$\rightarrow$A$\rightarrow$B second
order phase transition sequence. Note that the A-phase can be skipped,
since the B-phase is also accessible from the normal phase by
a second order phase transition.}
\end{table}

The gap function (\ref{gap}) in the  
A-phase,
\begin{equation}
\label{Agap}
\Delta_{\pm}({\bm k}) = |\eta_1|\left[a^2k_y^2+b^2k_x^2\right]^{1/2},
\end{equation}
is unitary (degenerate)  
and has 
cusp point nodes in the $[001]$ directions,  as shown in Fig.\ 1a. 
In the B-phase, the gap function  
is
\begin{eqnarray}
\Delta_{\pm}(\bm{k})& =& \bigg[ 
(|\eta_1|^2 b^2 + |\eta_2|^2 a^2)k_x^2 
+ |\eta_1|^2 a^2 k_y^2
+|\eta_2|^2 b^2 k_z^2  \nonumber \\ 
&& \pm 2 |\eta_1||\eta_2||k_x|\sqrt{a^2b^2k_x^2+a^4k_y^2+b^4k_z^2}
\bigg]^{1/2}.\label{Bgap}
\end{eqnarray}
In this case the gap function is non-unitary and degenerate
only where 
${\bm d}({\bm k}) \times {\bm d}^{*}({\bm k}) =0$, that is, along the
line $k_x = 0$. 
The gap has four nodes which are solutions to 
$\Delta_{-}(\bm{k})=0$. 
When $|\eta_1|^2b^2 > |\eta_2|^2a^2$ the nodes are found at
$k_y = 0$ and  $\sqrt{|\eta_1|^2b^2-|\eta_2|^2a^2}k_x = \pm |\eta_2| b k_z$,
shown in Figs.\ 1b)-1d),
and when  $|\eta_1|^2b^2 < |\eta_2|^2a^2$ they are found at
$k_z=0$ and $\sqrt{|\eta_2|^2a^2-|\eta_1|^2b^2}k_x = \pm |\eta_2| b k_y$.
A three dimensional rendering of the lower branch of the gap function
is shown in Fig.\ 2.
\begin{figure}[htb]
\epsfxsize=5.1in
\epsfbox[80 320 600 670]{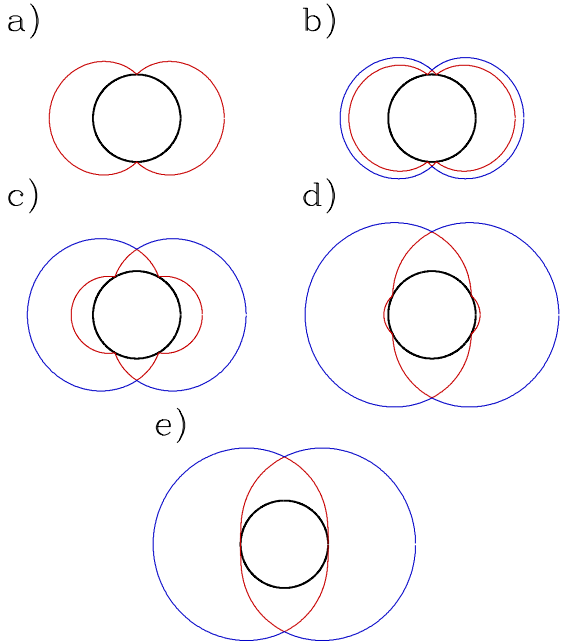}
\caption{\label{Bnodes} (Colour) The gap function $\Delta_{\pm}({\bm k})$
drawn over a spherical Fermi surface
(bold) in the $k_x$-$k_z$ plane.
In a) the gap function (red) is unitary and degenerate. In b)-e) the
gap function 
it is non-unitary and non-degenerate.  The lower branch  
$\Delta_{-}({\bm k})$ (red) and the upper branch
$\Delta_{+}({\bm k})$ (blue) are both shown.  
a) A-phase, $\eta_2=0$.  b) B-phase, $|\eta_2|a = 0.1 |\eta_1|b$.
c) B-phase, $|\eta_2|a = 0.5|\eta_1|b$. 
d) B-phase, $|\eta_2|a = 0.9|\eta_1|b$.
e) B-phase, $|\eta_2|a=|\eta_1|b$.}
\end{figure}
\begin{figure}[htb]
\resizebox{!}{8cm}{\includegraphics{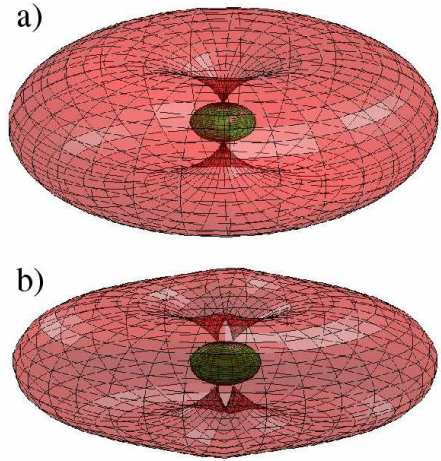}}
\caption{(Colour) The gap function drawn over a spherical Fermi surface for the
a) A-phase and b) B-phase.
In a) the gap function is unitary and degenerate. In b) the
gap function 
it is non-unitary and non-degenerate.  Only the lower branch of the gap function
$\Delta_{-}({\bm k})$ is shown.}  
\end{figure}

As discussed in the Introduction, the B-phase may evolve either
from the A-phase, with $|\eta_2|\ll |\eta_1|$, or directly from the normal phase, 
in which case $|\eta_2|\approx|\eta_1|$.
We now discuss these two scenarios in detail.

The order parameter of the A$\rightarrow$B transition is $\eta_2$,
which increases continuously from zero at the phase transition.
The two degenerate cusp nodes in the $[001]$ directions
in the A-phase (Fig.\ 1a)
split into four non-degenerate cusp nodes in the B-phase at
the phase transition (Fig.\ 1b). 

The order parameter of the normal$\rightarrow$B transition is 
$|\eta_1|=|\eta_2|$.
In this case, the B-phase resembles the $D_4(E)$ phase of
octahedral systems corresponding to 
the three-dimensional representations with components $(0,i,1)$.
In the Landau potential, the difference
between octahedral and tetrahedral appears only in 
sixth order and higher terms in the order parameter.\cite{Sergienko2004}   
Near the normal-to-superconducting phase transition, when all
components of the order parameter are small, the growth of the 
order parameter is governed by fourth order terms in the Landau potential,
which are identical for octahedral and tetrahedral systems,
so $|\eta_2|=|\eta_1|$ at the phase transition in both cases.
The difference between the gap functions of 
octahedral and tetrahedral systems
with 3D order parameter components $(0,i,1)$ 
is due to a difference in the basis functions (\ref{basis}): 
$|a|=|b|$ in octahedral 
systems.
Thus the  octahedral phase $(0,i,1)$
 has two non-degenerate smooth nodes in the 
$[100]$ directions  shown in Fig.\ 1e), while  
the tetrahedral system has four cusp nodes 
(Figs.\ 1b-1d).  

Thus the main difference between the two possible scenarios 
is the positioning
of the nodes at the onset of the B-phase.  
In the normal$\rightarrow$A$\rightarrow$B scenario, 
the nodes will always be found in pairs near the 
$[001]$ directions (Fig.\ 1b), while in the normal$\rightarrow$B scenario, 
the positions of the four
nodes are arbitrary (Fig.\ 1b-1d) and depend on the parameters $a$ and $b$.


\section{Density of states}
The low temperature form of the density of states (DOS) in
superconductors is governed 
by the  presence of nodes.\cite{Sigrist1991,Barash1996,Joynt2002}
In general, cusp-like point nodes give rise to a 
quadratic dependence on energy.
 
The DOS is given by\cite{Sigrist1991} 
\be
\label{dos}
N(\omega)=\frac{1}{(2\pi)^3}\int d^3k \sum_{\pm}\delta(\omega-E_{\pm}(\bm k)),
\ee     
where $E_{\pm}({\bm k}) = \sqrt{\varepsilon^2({\bm k})
+\Delta_{\pm}^2({\bm k})}$ and  $\varepsilon({\bm k})=\frac{k^2}{2m}-E_F$
is the
free particle energy.

\subsection{A-phase}
The gap function of the A-phase (\ref{Agap}) is 
unitary and non-degenerate (Fig.\ 1a). 
Since the main contributions to the integral come
from the vicinity of the nodes, the integral over ${\bm k}$ can be split into
two separate regions centred over each node, which are cut off such
that the total integrated region in $k$-space equals the 
Brillouin zone.\cite{Durst}
The nodes are degenerate and the contributions from each node are equal,
\begin{eqnarray}
\label{sum}
N(\omega) & =& \frac{4 v^2}{ab|\eta_1|^2(2\pi)^3}\int_0^{2\pi}d\phi \int_0^{\infty}dk_{\parallel}k_{\parallel} \nonumber \\
& & \int_{-\infty}^{\infty} dk_{\perp} 
\delta(\omega-E(k_{||},k_{\perp})) 
\end{eqnarray}
where $k_{\parallel}$ and $k_{\perp}$ are the momenta parallel and
perpendicular to the Fermi surface at the node, 
 $v^2 k_{||}^2 = |\eta_1|^2(a^2k_y^2 + b^2 k_x^2)$, $k_{\perp} = k_z - k_F$ and
$E(k_{||},k_{\perp}) \approx \sqrt{k_{\perp}^2 v_F^2 + k_{||}^2 v^2}$.
Changing variables again and using
$p_1=v_F k_{\perp}=p\,\cos\theta$, $p_2=v\,k_{\parallel}=p\,\sin\theta$,
we find
\begin{eqnarray}
N(\omega) & =&  \frac{4}{a b |\eta_1|^2(2\pi)^2 v_F}\int_0^{\infty}dp_2 p_2
\int_{-\infty}^{\infty}dp_1 \delta(\omega - E(p_1,p_2)) \nonumber  \\
& = & 
\frac{4}{a b |\eta_1|^2(2\pi)^2 v_F}
\int_0^{\pi}\sin\theta d\theta\int_{0}^{p_0}dp\,p^2
\delta(\omega-p) \\
& = & \frac{2\omega^2}{a b |\eta_1|^2 \pi^2 v_F}
\label{dosA}
\end{eqnarray}
where the cutoff $p_0$ is finally introduced in the last equation.
This result is equivalent to the usual result for a degenerate cusp node, 
$N(\omega) = \omega^2/ \pi^2 v_F v_g^2$,\cite{Joynt2002} apart from
a factor of 
two because there are two degenerate nodes in our
calculation.   In our case, the gap velocity, 
defined by ${\bm v}_g = {\bm \nabla}_{{\bm k}}\Delta({\bm k})$
is not the same in all directions
since the node is not rotationally symmetric, and so the geometric
average ${\bar v_g} = |\eta_1|(ab)^{1/2}$ appears.

Eq.\ \ref{dosA} is the density of states of the phase $(0,0,1)$ ($D_2(C_2)\times{\cal K}$)
at low temperatures.
However, according to the considerations 
outlined in Section I, this phase is identified as 
the A-phase, which is only found in a narrow region of phase space
just below $H_{c2}$.  Therefore, Eq.\ \ref{dosA} is not 
expected to be observed in 
PrOs$_4$Sb$_{12}$.  

\subsection{B-Phase}

In triplet, non-unitary phases, in general,
the gap function is non-degenerate, except
along some lines on the Fermi surface. 
All nodes are found in the lower energy branch of the gap
function $\Delta_{-}$, and the higher energy branch $\Delta_{+}$
is usually neglected.
However, if the nodes are found near the line where the gaps are degenerate
then both gaps should be taken into account.

To find the density of states in  the B-phase, 
we should consider the two different scenarios,
normal$\rightarrow$A-phase$\rightarrow$B-phase or
normal$\rightarrow$B-phase,
separately.
In the former scenario, $|\eta_2|\ll|\eta_1|$, and pairs of nodes 
are found on opposite sides of the Fermi surface.  The partners
in each pair are very close to each other and close to the 
gap degeneracy line, as shown in Fig.\ 1b).
In this case, the higher energy gap should not be neglected.
In the  normal$\rightarrow$B-phase scenario, the positions of the nodes
depend on the parameters $a$ and $b$ which are arbitrary.

\subsubsection{normal$\rightarrow$A-phase$\rightarrow$B-phase}
At the onset of the B-phase $|\eta_2|\ll|\eta_1|$,
and we will assume that $|\eta_1|^2b^2>|\eta_2|^2 a^2$.  Then the pairs of
nodes are found in the vicinity of $[001]$ in the plane
$k_y=0$, as shown in Fig.\ 1b).  The integration over
$k$-space is divided into four regions, which overlap
for nodes within a pair.  

The gap function in the vicinity of the nodes 
for the case when $|\eta_1|^2b^2 > |\eta_2|^2 a^2$
can be approximated by
\begin{equation}
\Delta({\bm k}) \approx 
\sqrt{|\eta_{1}|^2b^2-|\eta_2|^2a^2}\sqrt{k_{||}'^2+k_y'^2}
\end{equation}
where $k_y'=\frac{a}{b}\,k_y$ and
\begin{equation}
\label{k1}
k_{||}'=\frac{\sqrt{|\eta_1|^2b^2-|\eta_2|^2a^2}}{|\eta_1|b}k_x\pm\frac{|\eta_2|a}{|\eta_1|b}k_z.
\end{equation}
With this approximation, the `$-$' branch of the gap function continues 
smoothly to the `$+$' branch of the gap function at the line
where the gap function is degenerate.  Then two difficulties
are overcome at once:  both branches of the gap function are taken
into account, and the contributions
from each integration region are distinct, even though the regions
overlap.  
Each region yields the same contribution to the density of states,
\begin{widetext}
\begin{equation}
N(\omega) = \frac{4}{(2\pi)^3}\frac{b}{a}
\frac{v^2}{|\eta_1|^2b^2-|\eta_2|^2a^2}\int_0^{2\pi} d\phi
\int_0^{\infty}d k_{||} k_{||} \int_{-\infty}^{\infty}
dk_{\perp}\delta(\omega-E(k_{||},k_{\perp}))
\end{equation}
\end{widetext}
where
$v^2k_{||}^2 = 
(|\eta_{1}|^2b^2-|\eta_2|^2a^2)(k_{||}'^2+k_y'^2)$,
$k_{\perp} = \frac{\sqrt{|\eta_1|^2b^2-|\eta_2|^2a^2}}{|\eta_1|b}k_z\mp\frac{|\eta_2|a}{|\eta_1|b}k_x$
and $E(k_{||},k_{\perp}) \approx \sqrt{k_{\perp}^2v_F^2 + k_{||}^2 v^2}$
as before.
Then performing the same change of variables as in the A-phase 
calculation, we find
\begin{equation}
N(\omega) = \frac{b}{a}\frac{2\omega^2}{\pi^2v_F(|\eta_{1}|^2b^2-|\eta_2|^2a^2)}
\label{dos3}
\end{equation}   
Note that in the limit $|\eta_2|\rightarrow 0$
we recover the A-phase result, as expected.

\subsubsection{normal $\rightarrow$ B-phase}

In this situation, near the phase transition we have 
$|\eta_1|\approx |\eta_2|$, however the positions of the nodes depend
on the parameters $a$ and $b$, which are completely undetermined.  
Then there are three possibilities to consider.  The first is shown
in Fig. 1b), where  the nodes
appear in pairs such that the pairs are
close to the gap degeneracy line (if $|a|\ll |b|$ or $|b|\ll |a|$); in this
case the above calculation is valid and the result 
(\ref{dos3}) is obtained for $|\eta_1|\approx |\eta_2|$,
\begin{equation}
N(\omega) = \frac{b}{a}\frac{2 \omega^2}{\pi^2 v_F |\eta_1|^2 (b^2-a^2)}
\label{dos4}.
\end{equation}
Second, when all four nodes are spaced far apart as shown in Fig.\ 1c),
then the above calculations
are again valid and the result (\ref{dos4}) is obtained.  

Finally, the nodes may appear in pairs which are far away from the gap 
degeneracy line, as shown in Fig.\ 1d).  In this case the above treatment is
invalid. Here we have a crossover between 
$N(\omega) \sim \omega^2$ and $N(\omega) \sim |\omega|$,
which is the behaviour of the limiting case shown in Fig.\ 1e), {\em i.e.}, the
octahedral phase $(0,i,1)$, with smooth (quadratic) nodes.
Such behaviour is not observed in experiments, which could mean that
either the components of the order parameter are unequal (normal$\rightarrow$A-phase$\rightarrow$B-phase scenario)
or $a\neq b$. 

\section{Specific heat and nuclear spin relaxation rate}

The specific heat at low temperatures is given by\cite{Sigrist1991}
\be
\label{sh1}
C(T)=\frac{2}{T}\int_0^{\infty}d\omega\, \omega^2N(\omega)
\left[-\frac{\partial f}{\partial\omega}\right]
\ee
Eqs.\ \ref{dosA} and  \ref{dos3} yield
\be
\label{c1}
C(T)=\frac{14\,\pi^2}{15 v_F ab |\eta_1|^2}T^3
\ee 
for the A-phase, and
\be
\label{c2}
C(T)=\frac{b}{a}\frac{14\,\pi^2}{15 v_F (|\eta_1|^2b^2-|\eta_2|^2a^2)}T^3
\ee
for the B-phase.


The longitudinal nuclear spin-lattice relaxation rate is given by\cite{Sigrist1991}
\be
\label{l1}
\frac{(1/T_1)_{T}}{(1/T_1)_{T_c}}=2\frac{T}{T_c}\int_0^{\infty}d\omega\,N(\omega)N(\omega-\omega_0)
\left[-\frac{\partial f}{\partial\omega}\right].
\ee
In the limit of small nuclear resonance frequency $\omega_0$, one finds
\be
\label{l2}
\frac{(1/T_1)_{T}}{(1/T_1)_{T_c}}=\frac{28}{15\,\pi^4 v_F^2 a^2 b^2 |\eta_1|^4}\frac{T^5}{T_c}
\ee
in the A-phase,
while  in the B-phase it is
\be
\label{l3}
\frac{(1/T_1)_{T}}{(1/T_1)_{T_c}}=\frac{b^2}{a^2}\frac{28}{15\,\pi^4v_F^2
(|\eta_1|^2b^2-|\eta_2|^2a^2)^2}\frac{T^5}{T_c}.
\ee

These expressions give the low temperature behaviour of the
specific heat and nuclear relaxation rate in terms of the 
tetrahedral parameters $a$ and $b$ and the order parameter components
$\eta_1$ and $\eta_2$.

\section{Domains\label{sec-domains}}
Directional dependent measurements are the ideal way to observe
the anisotropy of the gap function.  However, such measurements may
be confounded by the presence of domains, different regions in space
where the components of the order parameter are interchanged.  
In this section we offer a brief discussion of domains 
for the A-phase and the B-phase.

The A-phase has three different domains
$(1,0,0)$, $(0,1,0)$ and $(0,0,1)$,
which, in the absence of unusual crystal shape or external fields,
are all expected to be present,
and will lead to the observation of the full tetrahedral symmetry.
Six (degenerate) nodes will be observed in the directions $\langle00\pm1\rangle$.
Now let us suppose that there is some kind of
external effect along the $z$-axis which effectively lowers the symmetry from 
$T_h$ to $D_{2h}$.  
In an octahedral system, either the single domain $(0,0,1)$,
with nodes in the 
$[00\pm1]$ directions will be favoured, {\em or}  the other two
domains, $(1,0,0)$ and $(0,1,0)$
will be favoured.  In the latter case, four nodes
would be observed in the directions $[\pm100]$ and $[0\pm 10]$.
However, because the crystal symmetry of \pr\ is tetrahedral
to begin with, any axial perturbation will lift the degeneracy of all
three domains, any of which could be favoured.
Therefore, in the A-phase, if all domains are present then tetrahedral symmetry
with six nodes will be observed.  
Otherwise, only one domain is present, the symmetry will be $D_2(C_2)$, with
two nodes.  It is not likely that two out of three domains would be present
in the A-phase, but could be possible if they were very close in energy.

The same arguments also hold for the more complicated B-phase.
Six domains are possible, with twenty-four non-degenerate 
nodes.  If there is a single
domain, then the symmetry is $D_2(E)$, and four nodes will be present.

\section{Conclusions}

In this article, we have attempted to give a physical
description 
and comparison of the sequences of phase transitions
normal$\rightarrow D_2(C_2)\times{\cal K}\rightarrow$~$D_2(E)$ and normal$\rightarrow$$D_2(E)$, which we identify with the phase transitions seen in experiments,
normal$\rightarrow$A$\rightarrow$B or normal$\rightarrow$B, respectively.
Although this description is derived entirely from basic 
considerations of 
symmetry, a complicated gap structure emerges
with several unusual features.
First, the positions of the nodes in the B-phase are not located
on any symmetry axes.  Although this is allowed by symmetry to occur
in crystals with other point groups, such a feature has never before been
considered.
Second, because the B-phase is triplet and non-unitary, there are
two non-degenerate gaps.  The only known example of this is
 Sr$_2$RuO$_4$, but in that
case the two gaps remain close in energy.\cite{Mackenzie2003}
In \pr, for a 
direct normal$\rightarrow$B transition, the energy difference is expected
to be large.  
Finally, the proposed A$\rightarrow$B transition, which is characterised by
the splitting into two of the degenerate nodes of the 
the A-phase, is highly unusual.  

In summary, we have proposed phase transition sequences in accordance
with experimental evidence available to date and studied its basic properties.
Superconductivity is best-described by a three component order 
parameter in the triplet channel.
The superconducting phase has  $D_2(E)$ symmetry, 
is non-unitary, and has four cusp nodes at unusual points on the 
Fermi surface.  
The presence of nodes leads to a quadratic dependence on
energy in the density of states, and power law behaviour in the 
specific heat and nuclear spin relaxation rate.  
There is also a second, higher energy, nodeless gap
which may be experimentally accessible.

\begin{acknowledgments}
We thank Ivan Sergienko for assistance
with the proof in the Appendix and
Ilya Vekhter for helpful discussions.  This work was
supported by NSERC of Canada.
\end{acknowledgments}

\appendix
\section{Proof of the existence of nodes in the 
$D_2(E)$ phase in the triplet channel}

In Section II, we found the gap function using 
basis functions given by (\ref{basis}), and order parameter
components 
$(0,i|\eta_2|,|\eta_1|)$.  
The gap function takes the form (\ref{Bgap}), which
vanishes either in the plane $k_y=0$ at the points defined by
$\sqrt{|\eta_1|^2b^2-|\eta_2|^2a^2}k_x = \pm |\eta_2| b k_z$ when
$|\eta_1|^2b^2 > |\eta_2|^2a^2$, or
in the plane $k_z=0$ at the points
$\sqrt{|\eta_2|^2a^2-|\eta_1|^2b^2}k_x = \pm |\eta_2| b k_y$ when
$|\eta_1|^2b^2 < |\eta_2|^2a^2$.
In Section II, only p-wave pairing (basis functions linear in ${\bm k}$)
was considered.  In order to rigorously demonstrate the 
existence of nodes all possible
higher order pairings must be included in the basis functions.
We now consider this most general case.

The most general form for the basis functions of the representation $T$
in $T_h$ is
\begin{eqnarray}
d_1 &= &(f(k_x,k_y,k_z), g(k_x,k_y,k_z), h(k_x,k_y,k_z)) \label{A1}\\
d_2 &= & (h(k_y,k_z,k_x), f(k_y,k_z,k_x), g(k_y,k_z,k_x)) \\
&  = & (h',f',g') \nonumber \\
d_3 & = & (g(k_z,k_x,k_y), h(k_z,k_x,k_y), f(k_z,k_x,k_y)
\label{A3} \\
& = & (g'',h'',f'') \nonumber 
\end{eqnarray}
where $f({\bm k})$ is odd in ${\bm k}$, $g(k_x,k_y,k_z)$ is odd in $k_z$ and
even in $k_x$ and $k_y$, and
$h(k_x,k_y,k_z)$ is odd in $k_y$ and even in $k_x$ and $k_z$.
Eventually, we will 
find 
solutions to $\Delta_{-}({\bm k}) = 0$
where one of the $k$'s is zero (in agreement with
the particular case of lowest order in $k$ basis functions
(\ref{basis})), so we set $f({\bm k})=0$ now.

Using (\ref{gap}), (\ref{d(k)}) and (\ref{A1}-\ref{A3}) one finds
\begin{eqnarray}
\Delta_{-}^2 &=& |\eta_1|^2 (g''^2+h''^2) + |\eta_2|^2(g'^2+h'^2)\nonumber \\
& & -2|\eta_1||\eta_2|\sqrt{h''^2g'^2+g''^2g'^2+h''^2h'^2}.
\end{eqnarray}
{\bf Case 1: $k_y=0$:} $g''$ vanishes and
\begin{equation}
\Delta_{-}^2 = (|\eta_1|h'' 
 -|\eta_2|\sqrt{g'^2+h'^2})^2.
\end{equation}
Nodes are found where $\Delta_{-}=0$, or  where the function
\begin{equation}
\phi_1(k_x,k_z) = h^2(k_z,k_x,0) - \frac{|\eta_2|^2}{|\eta_1|^2}
(g^2(0,k_z,k_x) + h^2(0,k_z,k_x))
\end{equation}
vanishes.\\
{\bf Case 2: $k_z=0$:} $h'$ vanishes and
\begin{equation}
\Delta_{-}^2 = (|\eta_2|g' 
 -|\eta_1|\sqrt{g''^2+h''^2})^2.
\end{equation}
Nodes are found where $\Delta_{-}=0$, or
where the function
\begin{equation}
\phi_2(k_x,k_y)  =   \frac{|\eta_2|^2}{|\eta_1|^2} g^2(k_y,0,k_x)
-(g^2(0,k_x,k_y) + h^2(0,k_x,k_y))
\end{equation} 
vanishes.

We have
\begin{eqnarray}
\phi_1(k_x,0) & = & h^2(0,k_x,0)
 - \frac{|\eta_2|^2}{|\eta_1|^2}
g^2(0,0,k_x)  \nonumber  \\
\phi_1(0,k_z) & = &  - \frac{|\eta_2|^2}{|\eta_1|^2} h^2(0,k_z,0) <0 \nonumber\\
\phi_2(k_x,0) & = &  \frac{|\eta_2|^2}{|\eta_1|^2} g^2(0,0,k_x) 
 -h^2(0,k_x,0)\nonumber \\
& =&  -\phi_1(k_x,0) \nonumber \\
\phi_2(0,k_y) & = & - g^2(0,0,k_y) <0 \nonumber
\end{eqnarray}
If $\phi_1(k_x,0) > 0$, then $\phi_1(k_x,k_z)$ changes sign, {\em i.e.},
there is a node of $\Delta_{-}$
in the $k_y=0$ plane somewhere between the positions $(k_x,0,0)$ and $(0,0,k_z)$. Symmetry requires
that there be (at least) four nodes  on the Fermi surface.
If $\phi_1(k_x,0) < 0$, then $\phi_2(k_x,k_y)$ changes sign, {\em i.e.}, there
are 
four nodes in
the $k_z=0$ plane.

Thus we have proved that, in general, the triplet phase with order
parameter components $(0,i|\eta_2|,|\eta_1|)$ has four nodes in either the
plane $k_y=0$ or $k_z=0$ at the positions 
$[\pm \alpha,0,\pm\beta]$ or $[\pm \alpha,\pm\beta,0]$, where
$\alpha$ and $\beta$ depend on the particular form of the basis functions.
These nodes are ``approximate", in the sense that they are a
consequence of symmetry and follow from 
the most general basis functions for the
$T$ representation.  These nodes are also 
``rigorous", since the state  $(0,i|\eta_2|,|\eta_1|)$ couples to no 
secondary superconducting order parameters.\cite{Sergienko2004}

\end{document}